\documentclass[12pt,preprint]{aastex}

\newcommand{\sqcm}{${\rm cm^{-2}}$}
\newcommand{\cubcm}{${\rm cm^{-3}}$}
\newcommand{\mum}{${\rm \mu m}$}
\newcommand{\kms}{${\rm km~s^{-1}}$} 
\newcommand{\waven}{${\rm cm^{-1}}$}
\newcommand{\thirteenco}{${\rm ^{13}CO}$}
\newcommand{\twelveco}{${\rm ^{12}CO}$}
\newcommand{\thirteencob}{${\bf ^{13}CO}$}
\newcommand{\twelvecob}{${\bf ^{12}CO}$}
\newcommand{\thirteencoi}{${\it ^{13}CO}$}
\newcommand{\twelvecoi}{${\it ^{12}CO}$}

\newcommand{\eighteenco}{${\rm C^{18}O}$}

\newcommand{\irsnine}{NGC 7538 : IRS9}

\usepackage{emulateapj5}

\slugcomment{To appear in ApJ 577 n1, 20 Sept. 2002 (submitted 6 April
2002; accepted 29 May 2002) }

\shorttitle{Detection of Solid $\bf ^{13}CO$ in Icy Grain Mantles}
\shortauthors{Boogert, Blake, \& Tielens}

\begin{document}

\title{High Resolution 4.7 \mum\ Keck/NIRSPEC Spectra of
Protostars.  II: Detection of the $\bf ^{13}CO$ Isotope in Icy Grain
Mantles\footnotemark}\footnotetext{The data presented herein were obtained at the
W.M. Keck Observatory, which is operated as a scientific partnership
among the California Institute of Technology, the University of
California and the National Aeronautics and Space Administration.  The
Observatory was made possible by the generous financial support of the
W.M. Keck Foundation.}

\author{A.C.A. Boogert\altaffilmark{2}, 
        G.A. Blake \altaffilmark{3},
        A.G.G.M. Tielens\altaffilmark{4,5}}

\altaffiltext{2}{California Institute of Technology,
              Department of Astronomy 105-24, Pasadena, CA 91125,
              USA; acab@astro.caltech.edu}
\altaffiltext{3}{California Institute of Technology, Division of
              Geological and Planetary Sciences 150-21, Pasadena, CA
              91125, USA}
\altaffiltext{4}{Kapteyn Astronomical Institute, P.O. Box 800, 9700 AV
              Groningen, the Netherlands}
\altaffiltext{5}{SRON, P.O. Box 800, 9700 AV Groningen, the
              Netherlands}

\begin{abstract}
The high resolution ($R$=25,000) infrared M band spectrum of the
massive protostar \irsnine\ shows a narrow absorption feature at 4.779
\mum\ (2092.3 \waven) which we attribute to the vibrational stretching
mode of the \thirteenco\ isotope in pure CO icy grain mantles.  This
is the first detection of \thirteenco\ in icy grain mantles in the
interstellar medium.  The \thirteenco\ band is a factor of 2.3
narrower than the apolar component of the \twelveco\ band. With this
in mind, we discuss the mechanisms that broaden solid state absorption
bands. It is shown that ellipsoidally shaped pure CO grains fit the
bands of both isotopes at the same time.  Slightly worse, but still
reasonable fits are also obtained by CO embedded in N$_2$--rich ices
and thermally processed O$_2$--rich ices. In addition, we report new
insights into the the nature and evolution of interstellar CO ices by
comparing the very high resolution multi-component solid \twelveco\
spectrum of \irsnine\ with that of the previously studied low mass
source L1489 IRS. The narrow absorption of apolar CO ices is present
in both spectra, but much stronger in \irsnine. It is superposed on a
smooth broad absorption feature well fitted by a combination of CO$_2$
and H$_2$O--rich laboratory CO ices. The abundances of the latter two
ices, scaled to the total H$_2$O ice column, are the same in both
sources. We thus suggest that thermal processing manifests itself as
evaporation of apolar ices only, and not the formation of CO$_2$ or
polar ices.  Finally, the decomposition of the \twelveco\ band is used
to derive the \twelveco/\thirteenco\ abundance ratio in apolar ices.
A ratio of \twelveco/\thirteenco=71$\pm$15 (3$\sigma$) is deduced, in
good agreement with gas phase CO studies ($\sim$77) and the solid
\twelveco$_2$/\thirteenco$_2$ ratio of 80$\pm$11 found in the same
line of sight. The implications for the chemical path along which
CO$_2$ is formed are discussed.
\end{abstract}


\section{Introduction}~\label{se13co:intro}

Ever since the detection of interstellar solid CO \citep{soif79,
lacy84} in the 4.67 \mum\ spectra of protostars and background
objects, the absorption band profile has been used as a diagnostic of
the composition and evolution of interstellar ices \citep{whit85,
sand88, tiel91, chia98, teix98}. The absorption band was found to
consist of a narrow feature accompanied by a broader feature at longer
wavelengths. With the help of laboratory simulations it was found that
the broad component is due to CO mixed with H$_2$O (`polar' ices) and
the narrow feature due to pure CO or CO mixed with apolar species such
as O$_2$, N$_2$ or CO$_2$. The relative depth of the apolar and polar
ices is thought to reflect thermal processing in the envelopes of
protostars, because these ices have quite different sublimation
temperatures (18 K versus 90 K respectively). Also, the knowledge
gained from observing the interstellar CO band is invaluable in
studying the outgassing behavior of cometary ices as comets approach
the sun \citep{saal88}.

We have therefore started a program to measure the interstellar CO ice
band at very high spectral resolution ($R=$25,000), more than an order
of magnitude higher than customary until now, using the NIRSPEC
spectrometer at the Keck II telescope. In Paper I of this program, we
presented the spectrum of the low mass protostar L1489 IRS in the
Taurus molecular cloud (Boogert, Hogerheijde, \& Blake 2002).  At the
high spectral resolution we discovered a new, third, CO component on
the short wavelength side of the absorption band, which is compatible
with absorption by CO$_2$--rich CO ices.  Combining the ice
observations with information obtained from the gas phase CO lines in
the same spectrum, we concluded that the CO ices are thermally
processed in the upper layers of the circumstellar disk surrounding
L1489 IRS.  In this Paper, we present the high resolution $M$ band
spectrum of the massive protostar \irsnine. This source has a rich and
well studied infrared ice band absorption spectrum \citep{whit96}.
The ices are thought to reside in a thick and young circumstellar
envelope, surrounding a modest hot core where the ices have evaporated
\citep{mitc90}.  Indeed, large gas phase depletion factors are needed
in envelope models explaining millimeter wave emission lines toward
\irsnine\ \citep{tak00}. The contribution of ice absorption from
unrelated cold foreground clouds in the NGC 7538 complex is likely
small given the weakness of ice bands toward nearby protostars with
more evolved envelopes (e.g. NGC 7538 : IRS1; \citealt{lacy84}).  The
CO ice band toward \irsnine\ has a large apolar component, tracing
unprocessed ices in the cold envelope \citep{sand88, tiel91, chia98},
and thus forms an interesting contrast with L1489 IRS. In addition to
the \twelveco\ band we report the first detection of solid
\thirteenco\ in \irsnine\ (and in the interstellar medium in general),
providing an independent tracer of the composition of apolar ices. It
also offers a reliable way of measuring the interstellar carbon
isotope ratio, in follow up to measurements of the solid
\twelveco$_2$/\thirteenco$_2$ ratio obtained with the {\it Infrared
Space Observatory} (ISO; \citealt{boog00}), and to gas phase
measurements (see \citealt{wils94}).

This Paper is structured as follows. The observations and data
reduction procedure are described in \S 2. The 2092 \waven\ absorption
band is identified with solid \thirteenco\ in \S 3.1.1 using
laboratory spectra. In \S 3.1.2 we compare the apolar component of the
\twelveco\ band with the \thirteenco\ band, explaining the factor 2.3
wider \twelveco\ band. After isolating the apolar component of the
\twelveco\ band from the underlying broader absorption, we derive the
interstellar solid \twelveco/\thirteenco\ abundance ratio in \S
3.2. The astrophysical implications of these results are discussed in
\S 4. A refined picture for the processing of CO ices is presented in
\S 4.1.  By comparing solid \twelveco/\thirteenco\ and
$^{12}$CO$_2$/$^{13}$CO$_2$ isotope ratios we discuss the chemical
pathway leading to the formation of interstellar CO$_2$ in \S 4.2.

\section{Observations}

The massive protostar \irsnine\ was observed with the NIRSPEC
spectrometer \citep{mcle98} at the Keck II telescope atop Mauna Kea on
UT 2001 August 8.  During the observations the sky was clear and dry,
and the seeing was good ($\sim 0.5''$ at 2.2 \mum).  NIRSPEC was used
in the echelle mode with the 0.43$\times 24''$ slit, providing a
resolving power of $R=\lambda/\Delta \lambda=25,000$ ($\sim$12 \kms)
with three Nyquist sampled settings covering the wavelength ranges
4.624--4.699 \mum, 4.684--4.756 \mum, and 4.754--4.822 \mum\ in the
atmospheric M transmission band.

The data were reduced in a standard way, using IDL routines. The
thermal background emission was removed by differencing the nodding
pair. Each nodding position was integrated on for 2 minutes, before
pointing the telescope to the other nodding position. A correction for
residual sky emission was performed by subtracting neighboring rows
from the stellar spectrum. The most critical step in the reduction of
this data is the correction for atmospheric absorption features. For
this purpose the standard star HR 8585 (A1V) was observed, which is
bright ($V$=3.78) and reasonably close to \irsnine\ (an airmass of
1.20 versus 1.34 for \irsnine). The spectral shape and hydrogen
absorption features in the standard were divided out with a Kurucz
model atmosphere.  After ratioing the standard and \irsnine\ spectra,
an overall good telluric correction was achieved, resulting in final
signal-to-noise values of respectively $\sim$40, 60, and 70 on the
unsmoothed data, for integration times of 6, 12, and 26 minutes on
each of the three settings.  The noise determination takes into
account small imperfections in the cancellation of the multitude of
weak telluric features, which we believe are the limiting factor in
the achievable signal-to-noise for this bright object (M$\sim$3 magn.;
\citealt{whit96}). Regions with strong atmospheric lines, i.e.  with
less than 50\% of the maximum transmission in each setting, leave
residuals and were removed from the final spectrum. This does not
affect the interstellar CO lines because at the time of the
observations these lines are shifted by as much as --82 \kms\ with
respect to the deep telluric CO lines.  The spectra were wavelength
calibrated on the atmospheric CO emission lines, and subsequently the
three settings were combined by applying relative multiplication
factors.  We have not attempted to flux calibrate the spectrum, since
we are interested in absorption features only.

The longest wavelength setting was also observed on two other nights
(UT 2001 August 7 and UT 2002 January 3). The presence, depth and
profile of the weak absorption feature of \thirteenco, the topic of
this paper (\S 3), are confirmed in these independent data. We do not
present these data here, because of their overall reduced quality due
to shorter integration time (2001 August) and poor system performance
(2002 January).

\section{Results}

\begin{figure*}[t!]
\includegraphics[angle=90, scale=0.80]{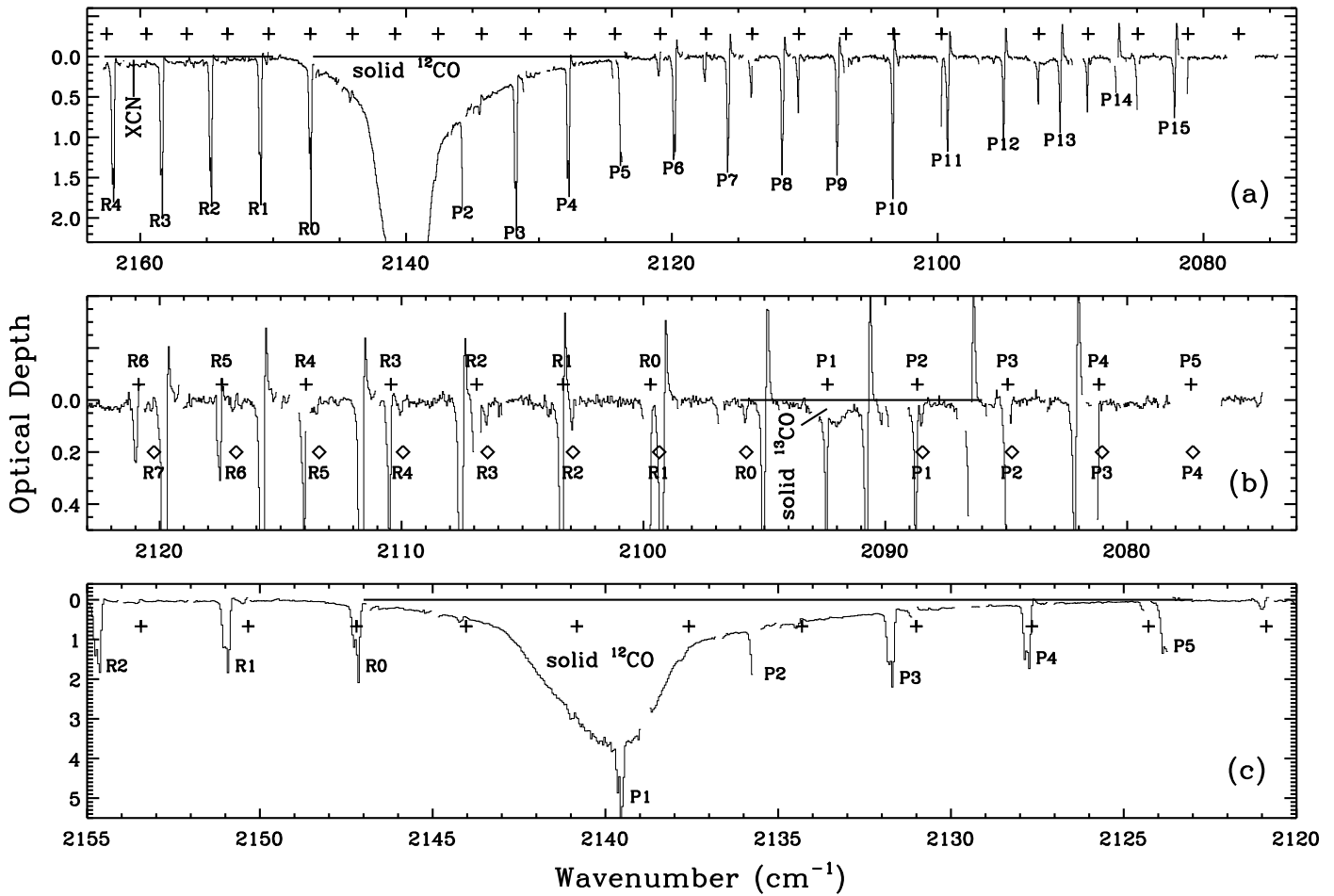}
\caption{Observed, unsmoothed, R=25,000 M band spectrum of \irsnine\
on an optical depth scale, with identifications of gas phase
\twelveco\ (deepest narrow lines), gas phase \thirteenco\ (`+' symbols
above spectrum), part of the shallow `XCN' band, and solid \twelveco\
(panel {\bf a}). A portion of the spectrum is enlarged in panel {\bf
b}, showing in detail the absorption band of solid \thirteenco\ at
2092 \waven\ along with the absorption lines of gas phase \thirteenco\
(`+' symbols) and \eighteenco\ (diamonds).  The emission component on
the red shifted side of the deep high $J$ level \twelveco\ absorption
lines are real. Panel {\bf c} shows in detail the deep absorption band
of solid \twelveco. The presence of the P(1) line in the trough of the
ice absorption band indicates that the ice band is not saturated. All
plotted spectra are corrected for object and earth velocity ($-$82
\kms). Wavelength regions with poor atmospheric transmission are not
plotted, causing the gaps in the spectra.}~\label{f:obs}
\end{figure*}

The most prominent feature in the $R$=25,000 spectrum of \irsnine\ is
absorption by solid \twelveco\ in the frequency range 2125--2150
\waven\ (Fig.~\ref{f:obs}). A much weaker and narrower absorption
feature is present at 2092 \waven, which is ascribed to the stretching
mode of solid \thirteenco\ (see \S 3.1.1). Throughout the full
2074--2164 \waven\ frequency range absorption lines originating from
the ro-vibrational transitions of \twelveco, \thirteenco, and
\eighteenco\ are present. The \twelveco\ lines also show an emission
component on the red shifted side, increasing in strength for higher
rotational $J$ levels. Some of the higher $J$ level \thirteenco\ lines
have emission components as well.  The gas phase features have
previously been analyzed in \citet{mitc90}.  Our study will therefore
focus on the analysis of the ice features.

We study the ice features on an optical depth scale. The continuum on
the long wavelength side of the \twelveco\ ice feature is smoothly
rising and can be fitted with a straight line. The apparent continuum
slope on the short wavelength side is, however, significantly
steeper. Most likely this is because of the presence of an additional
ice absorption band due to the stretching mode of a triply bonded
CN-bearing species, the so-called `XCN' band. Indeed, low resolution
spectral studies report the presence of this band toward \irsnine\
centered at 4.62 \mum\ \citep{lacy84, chia98, pend99}. We therefore
chose to extrapolate the straight line continuum defined at high
wavelength toward lower wavelengths. This results in an `XCN'
absorption feature with an optical depth of $\tau$=0.13 at 4.62 \mum.
This is less than the value of 0.31 quoted in \citet{pend99}, which we
believe is mainly ($\Delta \tau$=0.1) due to contamination by blended,
unresolved gas phase CO lines in their low resolution spectrum, but
also partly due to the lack of short wavelength continuum in our
spectrum.

Although little flux is left in the bottom of the solid \twelveco\
band, this band is not saturated as is shown by the presence of the
P(1) line of gaseous \twelveco\ in the bottom of the ice band
(Fig.\ref{f:obs}).  The peak optical depth of the solid \twelveco\
band is $\tau$(CO)=3.6$\pm$0.2, an important number for our isotope
ratio determination (\S 3.2). The uncertainty of 0.2 is determined
from three difference nodding pairs, and shows the data is reliable
even in extremely deep absorption bands.  However, $\tau$(CO) is
significantly larger than the value of 2.6 previously reported
\citep{lacy84, tiel91}. This is partly explained by instrumental
broadening of the narrow apolar CO band by the previous low resolution
spectrometers. We calculate that the depth of the CO band would be
reduced by $\Delta \tau$(CO)=0.6 and 0.3 at the resolution of the data
in \citet{lacy84} and \citet{tiel91} respectively. An additional
uncertainty may result from the continuum determination in the
presence of many unresolved strong gas phase CO absorption lines.

The newly detected absorption band of solid \thirteenco\ is an
independent tracer of the composition of icy grain mantles. Its
identification is discussed in \S 3.1.1. Extra information is obtained
by comparing the \thirteenco\ band with the apolar component of the
\twelveco\ band (\S 3.1.2).  In \S 3.2 we derive the
\twelveco/\thirteenco\ isotope ratio in interstellar ices, for which a
new approach to decompose the polar and apolar \twelveco\ ice
components is introduced.  The astrophysical implications of these
results are discussed in \S 4.

\subsection{New Constraints on the Composition of Apolar Ices}

\subsubsection{Identification of Interstellar Solid $^{13}CO$}

In order to characterize the absorption feature detected at 2092
\waven\ in the spectrum of \irsnine, Gaussian fits were carried
out. We find a peak frequency of $\nu$=2092.30$\pm$0.21 \waven, a peak
optical depth of $\tau$=0.089$\pm$0.010, and a width of
FWHM = 1.50$\pm$0.45 \waven\ (3$\sigma$ errors).  At the NIRSPEC
resolving power of $R=25,000$ ($\Delta \nu=0.084$ \waven) the feature
is well resolved, and fortunately well separated from the forest of
surrounding interstellar and telluric gas phase absorption and
emission features. Features providing the greatest interference are
the unresolved P(1) absorption line of interstellar gaseous
\thirteenco\ at 2092.39 \waven\ and a telluric feature at slightly
larger frequency. The deepest part of the telluric feature was removed
from the data. Despite these limitations, the fitted Gaussian
parameters are similar in the observations made on the three different
nights (\S 2).

The laboratory experiments on solid CO from the works of
\citet{ehre97} were used with the goal of identifying the 2092 \waven\
feature with absorption by solid \thirteenco\ and to further constrain
the composition of interstellar ices. These laboratory spectra have
the high spectral resolution (1.0 \waven) and signal-to-noise required
to study the band profile of the weak and narrow \thirteenco\
feature. Other relevant laboratory studies, such as the N$_2$:CO
experiments of Elsila, Allamandola, \& Sandford (1997) and the pure CO
study of \citet{bara98} unfortunately lack sufficient spectral
resolution to resolve and characterize the \thirteenco\ feature.

The laboratory profiles of the \thirteenco\ stretching mode were
analyzed in a way similar to that done for \twelveco\ \citep{boog02}.
After a careful baseline subtraction, the peak position and width were
determined as a function of ice composition and temperature.
Generally, the same trends found for the \twelveco\ band are
recognized in \thirteenco\ (Fig.~\ref{f:lab}). The band of
\thirteenco\ in a pure amorphous CO ice peaks at 2092.3 \waven\ and is
very narrow (FWHM=1.5 \waven).  An extreme broadening and shift to
larger wavelengths are observed when CO is mixed with H$_2$O ice,
because of the large dipole moment of the H$_2$O molecules.  Similarly
large effects are expected for other astrophysically relevant
molecules with large dipole moments (CH$_3$OH, NH$_3$), but no
laboratory experiments are available for these mixtures at present.
Mixtures of CO with the apolar species CO$_2$, and to a lesser degree
with O$_2$, can also give drastically broadened \thirteenco\ bands. A
fragile amorphous structure is formed between the CO molecules and
CO$_2$ and O$_2$, providing an absorption feature that is broadest at
mixing ratios of 1:1 \citep{ehre97}. Upon warming this structure is
destroyed and the band narrows, although in particular for CO$_2$
mixtures the width is still larger than that of a pure CO ice. In
contrast, only a very small broadening is observed in mixtures of CO
with N$_2$, even at 1:1 mixing ratios (FWHM=1.8 \waven).  Finally, the
peak position does not shift in mixtures of CO with O$_2$, but the
\thirteenco\ band (as well as the \twelveco\ band) shifts by 0.5--1
\waven\ to shorter wavelengths when mixed with CO$_2$ and by 0.5
\waven\ when mixed with N$_2$.

Using these results, we conclude that the peak and width of the 2092
\waven\ absorption feature observed toward \irsnine\ are best
explained by the stretching mode of \thirteenco\ in a pure CO ice
(Fig.~\ref{f:labfit1}b and d). At low laboratory temperatures, only
small amounts (less than 10\% in total) of H$_2$O, CO$_2$, and/or
O$_2$ are allowed in the ices. Larger concentrations would violate the
observed narrow width.  A laboratory CO spectrum with 16\% CO$_2$
content at a laboratory temperature of $T=$10 K can already be
excluded, for example (Fig.~\ref{f:labfit2}b).  At higher laboratory
temperatures, CO$_2$ mixtures still poorly match the observations. In
O$_2$ mixtures, the band narrows sufficiently at higher temperatures
($\sim$30 K) to give reasonable fits even at large O$_2$
concentrations ($>50$\%; Fig.~\ref{f:labfit2}d and f).  Finally, the
mixture N$_2$:CO=1:1 gives a somewhat poorer fit to the observed
\thirteenco\ band because of a $\sim$0.5 \waven\ shift to shorter
wavelengths (Fig.~\ref{f:labfit2}h). However, we consider this mixture
still reasonable because of the good fit to the \twelveco\ band (\S
3.1.2).

\vbox{
\begin{center}
\leavevmode 
\hbox{%
\epsfxsize\hsize 
\includegraphics[angle=90, scale=0.49]{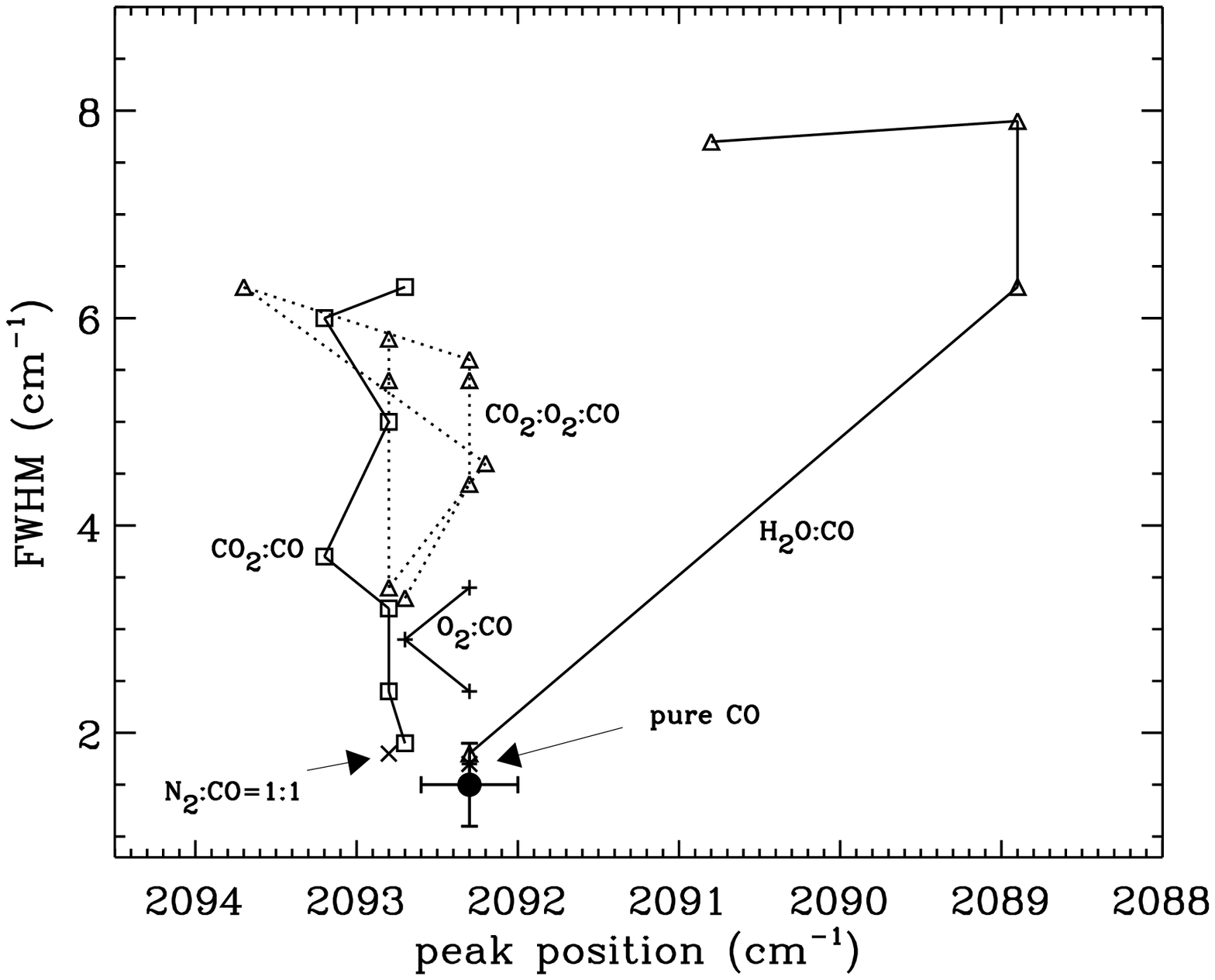}}
\figcaption{\footnotesize Diagram showing the effect of ice
composition on the peak position and width of the solid \thirteenco\
stretching mode, as determined from laboratory experiments.  The CO
concentration decreases upwards along each curve. In mixtures with
CO$_2$ and/or O$_2$ a maximum width is achieved at CO concentrations
of 50\%.  The width dependence is much smaller for N$_2$ mixtures. The
dot with the error bars represents the observed peak position and
width of the absorption feature detected toward \irsnine, which is
best fit by a pure CO ice, or by CO--rich ices with trace amounts
($\sim$10\%) of CO$_2$, O$_2$, and/or H$_2$O.~\label{f:lab}}
\end{center}}

\subsubsection{Comparison of \twelvecoi\ and \thirteencoi\ Band Profiles}

The profile of the solid \twelveco\ band toward \irsnine\ is dominated
by a prominent narrow absorption centered at 2140 \waven.  It shows
also absorption on an extended long wavelength wing, as well as in a
wing to the short wavelength side (Fig.~\ref{f:obs}).  Broad wing
absorption is not detected in the present signal-to-noise limited
spectrum of the \thirteenco\ band of \irsnine.  The origin of the
\twelveco\ wings and the decomposition from the central narrow
component is further discussed in \S 3.2.  Here we will concentrate on
a comparison of the \thirteenco\ absorption with the narrow 2140
\waven\ \twelveco\ component, which must, given their small width,
both originate from apolar ices.  The width of the \twelveco\ feature
(FWHM=3.5 \waven) is a factor of 2.3$\pm$0.7 (3$\sigma$) larger than
the width of the \thirteenco\ band.  What mechanism(s) can make the
interstellar \twelveco\ band broader than the \thirteenco\ band?

\begin{figure*}[t!]
\center
\includegraphics[angle=90, scale=0.50]{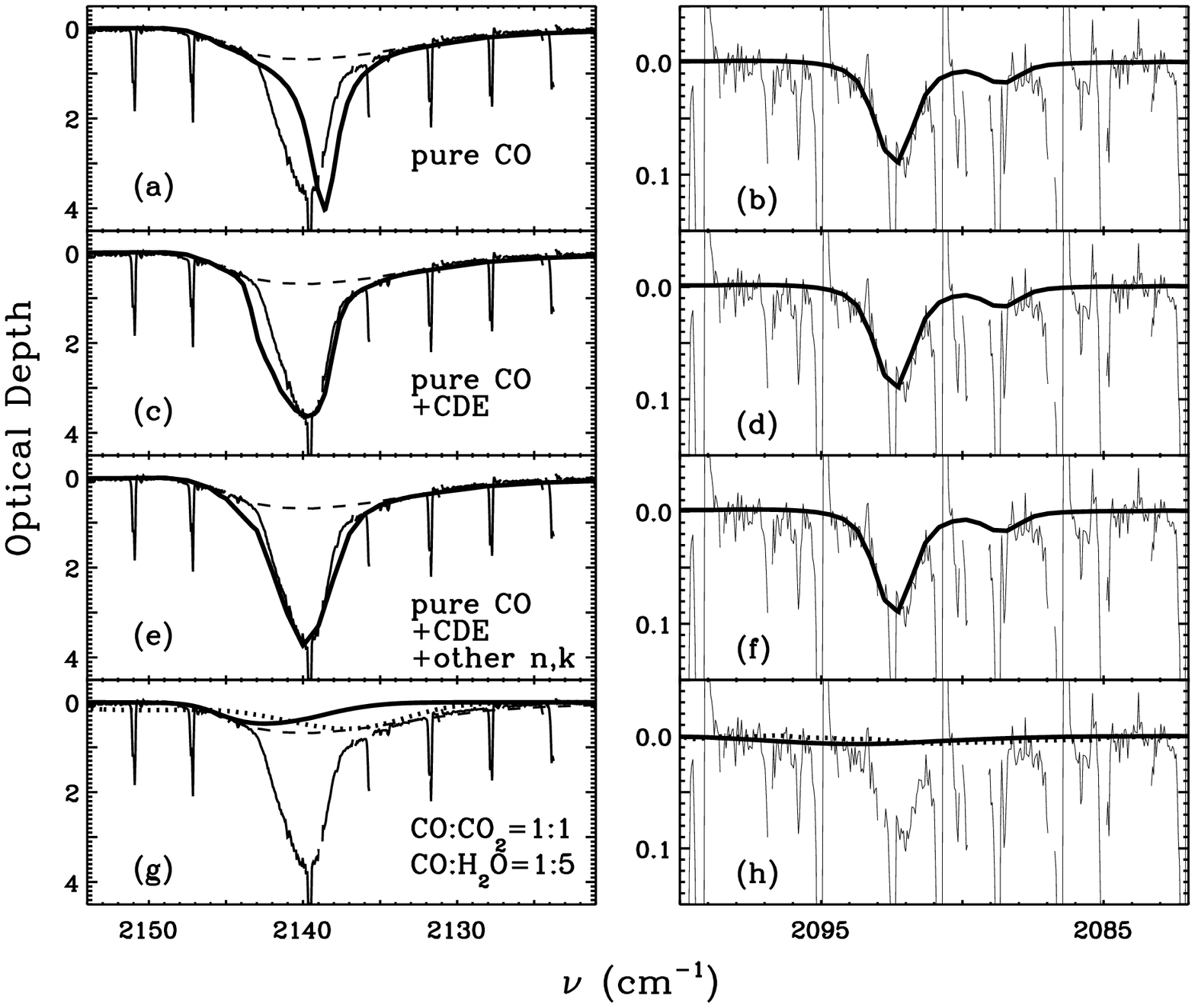}
\caption{Observed \twelveco\ (left panels) and \thirteenco\ (right
panels) absorption bands of \irsnine, fitted by laboratory and
calculated spectra of solid CO (thick lines; normalized to observed
peak optical depth). For the \twelveco\ band the shallow over-plotted
spectrum (dashed line) represents a fit to the long and short
wavelength wings of \irsnine\ obtained from the observed spectrum of
L1489 IRS (see Fig.~\ref{f:decomp}d).  Panels {\bf a} and {\bf b} show
added to this the pure CO ice spectrum observed in transmission in the
laboratory \citep{ehre97}, which clearly does not fit the \twelveco\
band. Panels {\bf c} and {\bf d} also show a pure CO ice, but now
assuming absorption by ellipsoidal grains (`CDE' model) providing a
much better fit. Panels {\bf e} and {\bf f} show the same grain model,
using the optical constants of \citet{bara98} for \twelveco\ (but
\thirteenco\ from \citealt{ehre97}).  Panels {\bf g} and {\bf h} show
a fit to the short and long wavelength wings of \twelveco\ by
CO:CO$_2$=1:1 (thick solid line) and CO:H$_2$O=1:5 (dots) mixtures
respectively, which have a negligible contribution to the \thirteenco\
absorption band.}~\label{f:labfit1}

\includegraphics[angle=90, scale=0.50]{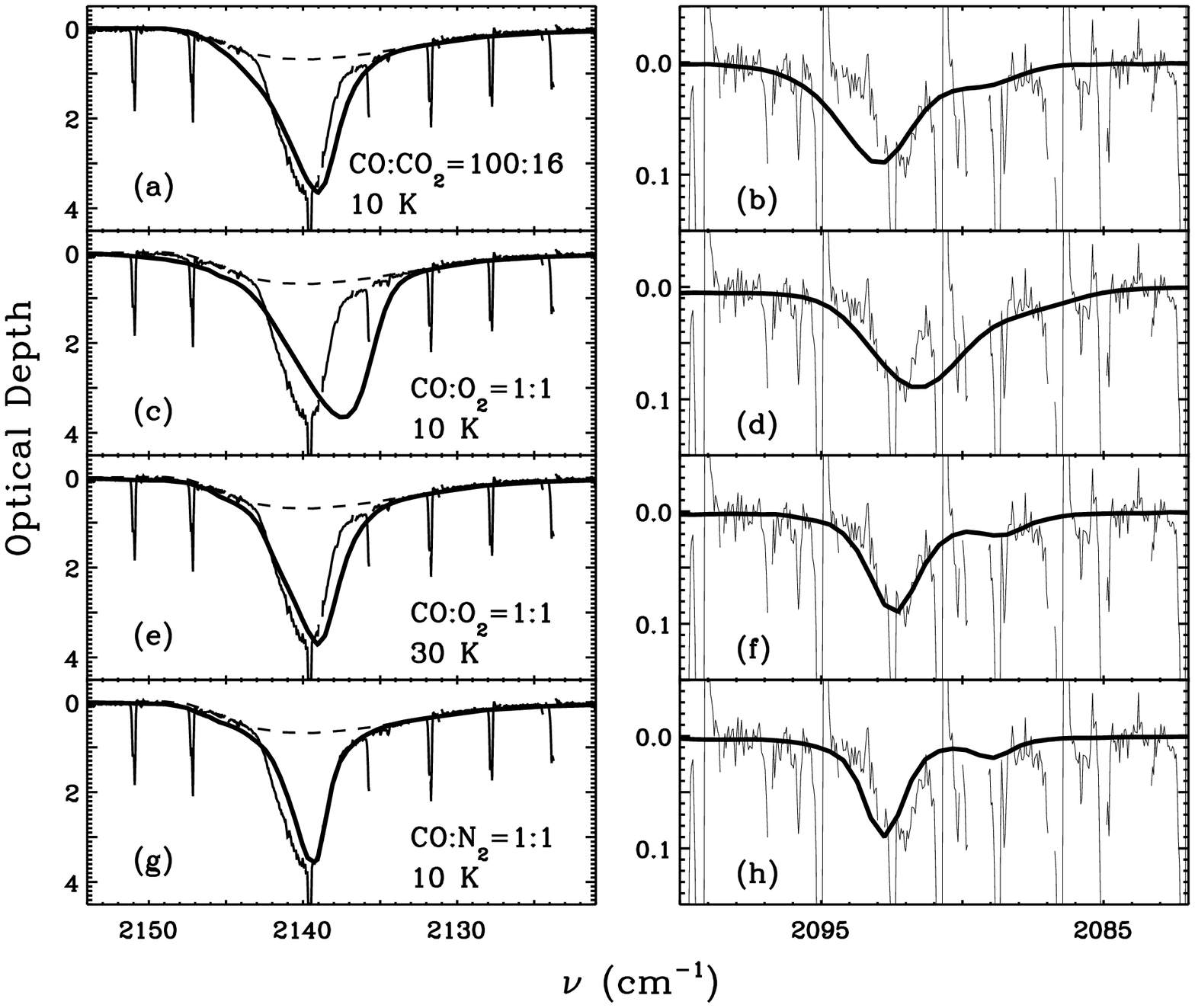}
\caption{Same as for Fig.~\ref{f:labfit1}, but now for comparing the
apolar component in \irsnine\ with various laboratory mixtures.
Panels {\bf a} and {\bf b} show the mixture CO:CO$_2$=100:16 as
observed in transmission in the laboratory. Panels {\bf c} and {\bf d}
show the mixture CO:O$_2$=1:1 at $T=$10 K, while panels {\bf e} and
{\bf f} show the same mixture at $T=$30 K.  Panels {\bf g} and {\bf h}
show CO:N$_2$=1:1.  The effect of particle shape corrections is small
for all mixtures, except for the \twelveco\ band in CO:CO$_2$=100:16.
We conclude that all these mixtures give poorer fits to the
observations compared to pure CO spectra (Fig.~\ref{f:labfit1}), but
CO:O$_2$=1:1 (provided it is thermally processed) and CO:N$_2$=1:1 are
still considered reasonable.}~\label{f:labfit2}
\end{figure*}

The width of absorption bands of molecules embedded in solid matrices
is affected by a number of difficult to disentangle mechanisms. In
experiments of CO isolated in annealed N$_2$ matrices, the width of
the CO bands was found to depend strongly on the CO/N$_2$ ratio
(Dubost, Charneau, \& Harig 1982).  This `concentration broadening' is
the result of shifts in the transition frequency because of static and
dynamic interactions between the CO molecules themselves. The static
interaction is due to the electric field generated by the permanent
dipole moment of one CO molecule working on other nearby CO molecules
or due to the electric field created by crystalline defects. The
dynamic interaction is due to the resonance of oscillating electric
fields generated by different CO molecules.  The frequency shifts
induced in both cases result in broadening of the absorption band
because of the distribution of the local field generated by the
different configurations of the neighboring molecules.  The smallness
of the permanent dipole moment of CO makes the broadening from dynamic
interaction significantly larger than broadening induced by static
interaction. This greatly enhances the width of the \twelveco\ band
with respect to the \thirteenco\ band, because the concentration of
\twelveco\ molecules is a factor of 90 higher. There is no dynamic
interaction of \thirteenco\ with \twelveco\ molecules, because their
transition frequencies are several tens of wavenumbers apart.  Thus,
at a mixing ratio of CO:N$_2$=1:100, the band of \twelveco\ was found
to be a factor of 5 wider than the \thirteenco\ band \citep{dubo82}.
However, at such a low CO concentration the bands of both isotopes are
an order of magnitude too narrow compared to the bands observed toward
\irsnine. At higher CO concentrations, the width of both bands
increases, but that of the \twelveco\ band grows more slowly because
the dynamic and static broadening mechanisms are not additive, and in
fact the static fields detune the dynamic interactions. As a result, a
pure, annealed CO ice has an FWHM=1.7 \waven\ for \twelveco, and
FWHM=1.1 \waven\ for \thirteenco\ \citep{ewin61, dubo82}.  These
widths increase by $\sim$0.5 \waven\ when the CO ice structure is
amorphous \citep{sand88}. Thus, a pure CO ice that may be amorphous
provides a good fit to the \thirteenco\ band (FWHM=1.5 \waven)
observed toward \irsnine, but this pure CO ice has a \twelveco\ band
that is still a factor of 1.60 too narrow.

An additional mechanism is therefore needed that can selectively
broaden the \twelveco\ band. A viable mechanism is interaction of
light with CO--rich interstellar icy grains.  Electromagnetic
radiation can polarize the \twelveco\ dipoles, which induces electric
fields within the grain that oscillate at the \twelveco\ vibration
transition frequency (e.g. \citealt{bohr83}).  This creates resonances
with the dipole electric field, resulting in frequency shifts and band
broadening, but only at sufficiently high CO concentrations ($>$30\%;
\citealt{tiel91}). Therefore, the profile and position of the band of
the diluted \thirteenco\ molecules are not affected by these particle
shape effects.  Using the calculations in the small particle limit
presented in \citet{ehre97}, we thus find that acceptable fits to the
\twelveco\ band are obtained for ellipsoidally shaped particles, in
particular with a continuous distribution of shapes (`CDE';
Fig.~\ref{f:labfit1}c). Although there is a mismatch on the short
wavelength side of the 2140 \waven\ feature, this can be accounted for
by assuming other particle shapes or by uncertainties in the optical
constants of CO. A better fit is obtained (Fig.~\ref{f:labfit1}e) if
we take the slightly different optical constants of \citet{bara98}.
Note that the shape of the \thirteenco\ band indeed remains unchanged
for different particle shapes (Fig.~\ref{f:labfit1}b and d).

Finally, an alternative way to preferentially broaden the \twelveco\
band is to change the ice composition. In the database of
\citet{ehre97} the ratio of \twelveco\ to \thirteenco\ band width is
$\sim$1.3 for most mixtures. At low laboratory temperatures ($T=10$ K)
the mixture N$_2$:CO=1:1 has the largest width ratio (1.5), while at
high laboratory temperatures ($T=$ 30 K) mixtures with O$_2$/CO$\geq$1
have width ratios of up to 1.8. The N$_2$ mixture provides the best
fit to the \twelveco\ band, and the heated O$_2$ mixture fits the
\thirteenco\ band better (\S 3.1.1; Fig.~\ref{f:labfit2}e--h). We thus
consider these mixtures as reasonable, but not as good as pure CO.

To conclude, the peak position and width of the \thirteenco\ band
observed in the direction of \irsnine\ are well explained by a pure CO
ice that might be amorphous. To fit the \twelveco\ band with the same
pure ice a mechanism is needed that preferentially broadens the
\twelveco\ band. Interaction of light with small particles has this
effect. It also shifts the \twelveco\ band to the observed peak
position.  Alternatively, the observed band widths and peak positions
are reasonably matched with N$_2$:CO=1:1 ices or thermally processed
O$_2$--rich ices.  Whether the {\it required} mild degree of thermal
processing of O$_2$ ices is realistic in the \irsnine\ line of sight
is further discussed in \S 4.1.

\subsection{Deriving the \twelvecob/\thirteencob\ Isotope 
Ratio in Interstellar Ices}

In order to determine an accurate \twelveco/\thirteenco\ ratio it is
necessary to correct the \twelveco\ absorption band for absorption in
its extended wings.  This requires knowledge of the intrinsic shape of
the absorption features responsible for the apparent wings.  Insight
is gained by comparing the observed \twelveco\ profile of \irsnine\
with that of the low mass protostar L1489 IRS \citep{boog02}. The two
spectra have similarly high signal-to-noise values at the same high
spectral resolution, allowing a detailed comparison. The absorption in
the wings is much more prominent in L1489 IRS with respect to that of
the central 2140 \waven\ peak (Fig.~\ref{f:decomp}a). It is found that
the \twelveco\ profile of \irsnine\ can be excellently reproduced by
multiplying the L1489 IRS spectrum with a factor of 1.5 to match the
wings (Fig.~\ref{f:decomp}b), and adding a Gaussian with peak
$\nu=$2139.8 \waven, width FWHM=3.5 \waven\ (\S 3.1.2) and an optical
depth of 2.4 (Fig.~\ref{f:decomp}c). This implies that the central
2140 \waven\ absorption and the wings are spectroscopically
independent CO ice components.  We can now determine the intrinsic
shape of the absorption in the wings by {\it subtracting} the same
Gaussian ---but different optical depth--- from the observed L1489 IRS
profile. The main uncertainty here is the depth of the Gaussian. Two
extreme cases can be distinguished: a relatively small correction of
$\tau=0.4$ (corresponding to 0.6 when scaled to the wings of \irsnine)
such that a smooth profile remains, or a factor of two larger
correction such that the optical depth at 2140 \waven\ is zero and two
separate long 
\vbox{
\begin{center}
\leavevmode 
\hbox{%
\epsfxsize\hsize 
\includegraphics[angle=90, scale=0.45]{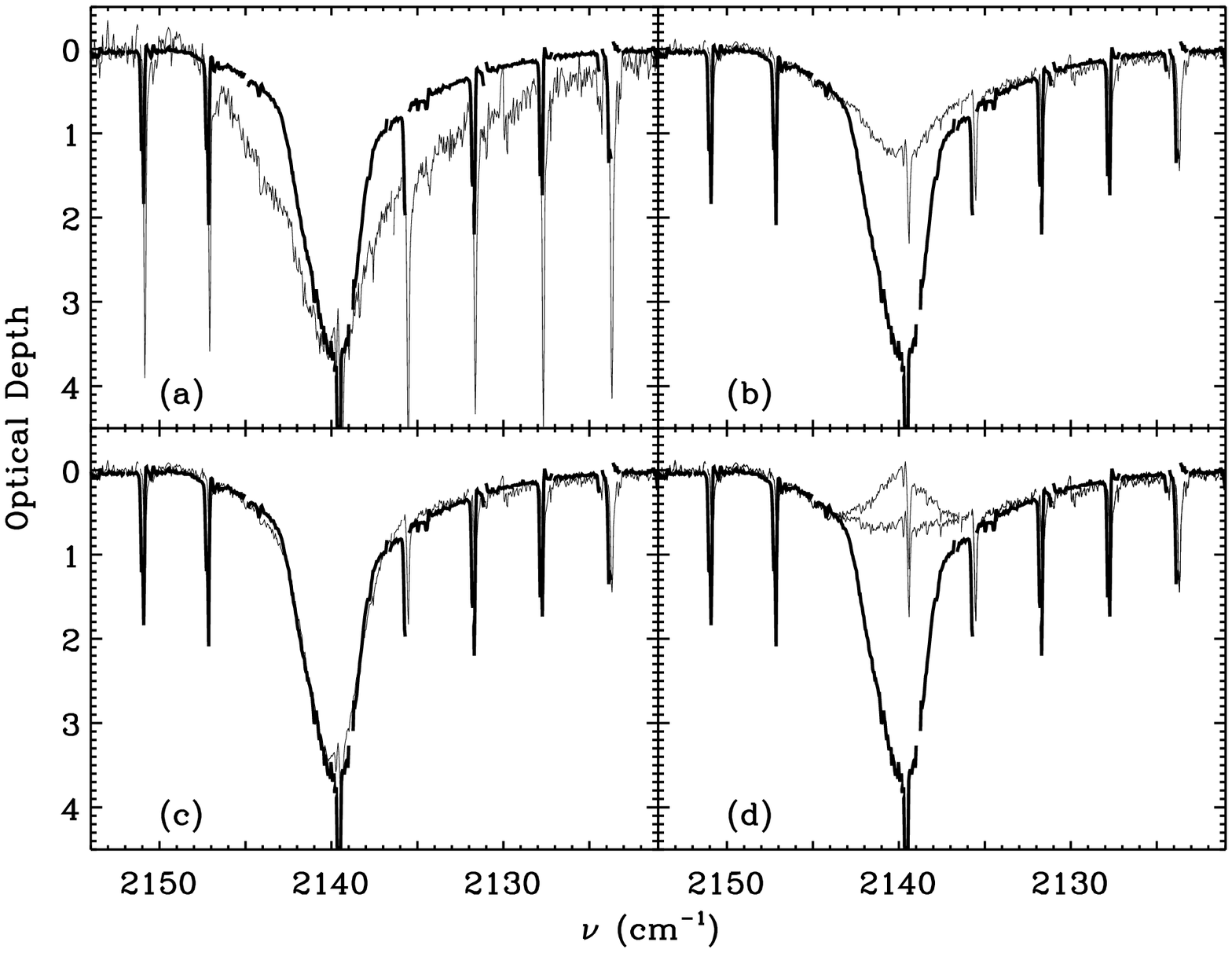}}
\figcaption{\footnotesize Comparison of the \twelveco\ spectrum of
\irsnine\ (thick line in each panel) to that of the low mass protostar
L1489 IRS (thin lines; \citealt{boog02}), providing insight into the
nature of this band. In panel {\bf a} the L1489 IRS spectrum is scaled
to the peak optical depth of the \irsnine\ spectrum by multiplication
with a factor of 4.3. In panel {\bf b} the wings of L1489 IRS are
scaled to the wings of \irsnine, by multiplication with a factor of
1.5.  In panel {\bf c} the \irsnine\ spectrum is excellently
reproduced by adding to the scaled L1489 IRS spectrum a Gaussian with
a peak optical depth of $\tau$=2.4. In panel {\bf d} the central
narrow peak in L1489 IRS is corrected for in two different ways by
subtracting the same Gaussian with $\tau$=0.6 and $\tau$=1.2
respectively from L1489 IRS (see text for more details).  In each
panel, the narrow absorption lines superposed on the ice spectra are
due to gas phase CO in the same line of sight.~\label{f:decomp}}
\end{center}}
and short wavelength absorption features remain
(Fig.~\ref{f:decomp}d).  We argue that the former case is most likely,
because with the presently available database of laboratory spectra it
is impossible to obtain two separate absorption features.  As
discussed in \citet{boog02}, the long wavelength wing of the
\twelveco\ band is very likely due to CO diluted in H$_2$O--rich ices
(see also \citealt{tiel91, chia98}), while the short wavelength wing
is well fitted with a CO:CO$_2$ mixture.  Together, they form the
smooth broad absorption feature underlying the narrow 2140 \waven\
band (however, see \S 4.1 for an alternative interpretation). Using
these laboratory mixtures, and assuming a nominal isotope ratio (see
below), it can be seen that the \thirteenco\ band in these ices is
much broader than the observed 2092 \waven\ feature and that the peak
optical depth is negligible (Fig.~\ref{f:labfit1}g and h).

For the column density and isotope ratio determinations that follow we
will therefore assume that the narrow 2140 \waven\ component is
superposed on a smooth underlying broad absorption feature, the shape
of which we determine from the L1489 IRS spectrum as described
above. We thus find that the integrated area of the narrow 2140
\waven\ feature toward \irsnine\ is 10.4 \waven, which is about half
the total integrated area (20.2 \waven).  If the second correction
method were applicable, the integrated area for the 2140 \waven\
feature would be 11\% larger.

The column density of ices is derived by dividing the integrated
optical depth over the laboratory measured intrinsic band strength
$A$.  The band strength\footnote{we will adopt the notation $A^{12}$
and $A^{13}$ for the band strengths of the stretching modes of
\twelveco\ and \thirteenco\ respectively} of \twelveco\ was measured
to be $A^{12} = 1.1\times 10^{-17}$ cm molecule$^{-1}$ \citep{jian75}.
The absolute value of $A^{13}$ has not been measured, but its value
relative to $A^{12}$ can be determined by measuring the integrated
optical depths of the \thirteenco\ and \twelveco\ stretching modes in
laboratory experiments. Using the pure CO spectrum of \citet{ehre97},
and taking into account a laboratory isotope ratio of 89, we thus find
that $A^{12}$ and $A^{13}$ are identical. This contradicts the study
of \citet{gera95} which reports that $A^{13}$ is 18\% larger than
$A^{12}$. It is likely that \citet{gera95} included a close-by
satellite band at 2088.6 \waven\ (Fig.~\ref{f:labfit1}) in the
integrated optical depth of \thirteenco\ (P. Gerakines, private
communication). This satellite band, however, must originate from
solid \eighteenco, given its position and strength
\citep{ewin61}. Finally, we find that $A^{12}$, but not $A^{13}$,
increases with 10\% when the spectrum is corrected for absorption by
small particles (as was also found by \citealt{tiel91}).  In
interstellar space this is a realistic case (\S 4.1), and we therefore
conclude that the interstellar isotopic \twelveco/\thirteenco\ column
density ratio is given by
\begin{equation}
\frac{\rm ^{12}CO}{\rm ^{13}CO}=\frac{\tau_{\rm int}^{12}\times
A^{13}}{\tau_{\rm int}^{13}\times A^{12}\times 1.10}=\frac{\tau_{\rm
int}^{12}}{\tau_{\rm int}^{13}\times 1.10}{\rm ,}
\end{equation}
where $\tau_{\rm int}^{12}$ and $\tau_{\rm int}^{13}$ are the
integrated optical depths of the observed \twelveco\ and \thirteenco\
bands. Deriving $\tau_{\rm int}^{13}$ from the Gaussian fit in \S
3.1.1, and $\tau_{\rm int}^{12}$ with the method described above, we
find
\begin{equation}
\frac{\rm ^{12}CO}{\rm ^{13}CO}=71\pm 15 (3\sigma).
\end{equation}
The $3\sigma$ uncertainty reflects the uncertainty in the peak depth
of the \twelveco\ (because of the low signal in the bottom of the
feature; \S 3) and \thirteenco\ features.  The 30\% error in the
observed FWHM of \thirteenco\ can be neglected if one assumes that a
pure CO ice is responsible for the \thirteenco\ absorption (FWHM=1.50
\waven).  A broadening by small amounts of CO$_2$, H$_2$O or O$_2$ is
allowed by the uncertainty in the width of the feature (\S
3.1.1). This would reduce the isotope ratio by 30\% (3$\sigma$),
i.e. such broadening imposes a lower limit of \twelveco/\thirteenco$>
50$.  Another systematic error is the extraction of the 2140 \waven\
component from the observed \twelveco\ profile, which could increase
the ratio by at most 11\%.

\section{Astrophysical Implications, Summary and Future Work}

\subsection{The Nature and Evolution of CO ices}

It has been claimed that CO ices evolve by the heating effects of
nearby protostars \citep{tiel91, chia98}.  Sources in different
evolutionary stages may thus have ices of different composition or
structure in their neighborhood. It is also expected that gradients
in the physical conditions result in different ices as a function of
distance to the protostar. In particular, the abundance ratio of
volatile apolar CO ices versus less volatile polar CO ices may reflect
the thermal distribution of the dust along the line of sight, much
like the structure of polar CO$_2$-containing ices traces the thermal
evolution of massive protostellar envelopes \citep{gera99, boog00}.
This might be accompanied by energetic processing of CO ices,
resulting in the formation of new species, such as CO$_2$ and perhaps
`XCN' close to the star, although this is less well established.

\begin{table*}[t!]
\center
\caption{Column Densities of Ices}~\label{t:colden}
\begin{tabular}{lccc}
\noalign{\smallskip} 
\noalign{\smallskip} 
\tableline
\noalign{\smallskip} 
Species	 &  \multicolumn{2}{c}{$N/N$(H$_2$O)$\times$100\%}   & Reference \\
       	 &  \irsnine  & L1489 IRS        &           \\
\noalign{\smallskip}  
\tableline
\noalign{\smallskip} 
H$_2$O	                &  100 & 100 & 1, 2 \\
CO$_2$	                &  25  & 17  & 3, 4 \\
CO apolar               &  15  & 3   & 5, 6 \\
CO long $\lambda$  wing &   8  & 7   & 5, 6 \\
CO short $\lambda$ wing &   4  & 3   & 5, 6 \\
\noalign{\smallskip} 
\tableline
\multicolumn{4}{p{11cm}}{References:
(1) $N$(H$_2$O)=64$\times 10^{17}$ \sqcm\ from \citealt{alla92};
(2) $N$(H$_2$O)=47$\times 10^{17}$ \sqcm\  from \citealt{sato90};
(3) \citealt{gera99};
(4) not observed, but assuming the typical value from \citealt{gera99}; 
(5) this work; 
(6) \citealt{boog02}}
\end{tabular}
\end{table*}

With the high spectral resolution and high signal-to-noise
observations of solid \twelveco\ and \thirteenco\ presented in this
paper, a refined picture of the evolution of interstellar CO ices can
be constructed. The \twelveco\ spectra of \irsnine\ and L1489 IRS (\S
3.2) can be taken as templates for unprocessed and processed ices
respectively (group I and II spectra in the nomenclature of
\citealt{chia98}). In \S 3.2 it was shown that the unprocessed
spectrum of \irsnine\ can be reproduced by simply adding a narrow
Gaussian (representing apolar ices at 2140 \waven) to the processed
spectrum of L1489 IRS.  Thus, processing manifests itself largely as
evaporation of the most volatile ices: the abundance of apolar ices
normalized to the H$_2$O ice column is a factor of 5 lower in the warm
upper layers of the L1489 IRS circumstellar disk compared to the cold
envelope of \irsnine\ (Table~\ref{t:colden}). In contrast, the
composition, structure, and abundances of the ices responsible for the
short and long wavelength wings remain unchanged.  Both sources have
abundances of 3.5\% and 7.5\% for the CO ices responsible for these
wings respectively. If we assume that the wing at short wavelengths is
caused by CO$_2$--rich CO ices\footnote{At present there are no
laboratory experiments supporting the idea that the same carrier is
responsible for both wings.} (\citealt{boog02}; consistent with the
high CO$_2$/CO column density ratio; Table~\ref{t:colden}), then the
similar abundance for both sources indicates that CO ice processing
does not produce significant quantities of CO$_2$ molecules.  This is
sustained by the fact that the position and width of the apolar 2140
\waven\ feature is the same for both objects.  The weakness of
energetic processing in the presence of strong thermal processing in
the L1489 IRS line of sight is also evident by the absence of a strong
2165 \waven\ (4.62 \mum) XCN band (we estimate $\tau$[XCN]$\leq$ 0.14
taking alternative continua compared to \citealt{boog02}).  The
non-correlation of energetically processed ices and (un)processed
apolar ices along one line of sight seems to be a common
characteristic of interstellar ices \citep{pend99, whit01} and may
well reflect a spatial segregation associated with protostellar
activity.  Finally, the comparison between these different lines of
sight does not favor the model in which CO molecules migrate into an
underlying porous H$_2$O ice layer at increased temperatures, because
this would result in higher abundances of CO ices responsible for the
long wavelength, polar wing \citep{thi02}. Of course, these hypotheses
must be tested against future high resolution observations of a larger
sample of protostars.

Additional constraints on the nature of the apolar CO component are
obtained by comparing the \twelveco\ and \thirteenco\ band widths and
peak positions. The \thirteenco\ band is best explained by a pure,
perhaps amorphous, CO ice.  The apolar component of the \twelveco\
band is a factor of 2.3 broader. This may well be caused by
interaction of light with randomly oriented CO--rich particles that
have a distribution of ellipsoidally shapes (CDE), which broadens the
\twelveco\ band but leaves the \thirteenco\ band
unchanged. Interestingly, ellipsoidally shaped grains were also found
to give best fits to the interstellar CO$_2$ bands \citep{gera99}. The
CDE distribution mimics irregularly shaped ice grains \citep{tiel91}
or grain clustering \citep{roul91}.

The case that pure CO ices are responsible for the apolar component
would imply that at the time of the ice mantle formation the abundance
of CO is larger than that of the apolar species O$_2$ and N$_2$.  It
has been argued that pure interstellar CO ices may not be realistic
given the large cosmic abundances of O and N \citep{elsi97}.  The
amount of O and N locked up in O$_2$ and N$_2$ is, however, poorly
constrained.  Direct detection of these species in the solid state by
their vibrational bands is complicated, and only high upper limits
have been derived \citep{vand99, sand01}.  Nevertheless, N$_2$ and
thermally processed O$_2$--rich CO ices give only slightly poorer fits
to the \thirteenco\ and \twelveco\ apolar component (\S 3.1.2; see
also discussions in \citet{chia98} and \citet{vand99} for \twelveco).
The laboratory temperature required for the O$_2$--rich ices to fit
the width of the bands of both isotopes is 30 K, close to the
sublimation temperature (Fig.~\ref{f:labfit2}c--f). At the low vapor
pressures and large time scales in interstellar space these
temperatures scale down to $\sim$18 K \citep{naka80}. While it is
likely that these conditions occur somewhere in the envelope or
\irsnine, one may question how at these temperatures so close to the
sublimation temperature the abundance of apolar ices can be so large.
Concluding, although pure CO ices are spectroscopically slightly
preferred, from an astrophysical point of view both pure CO, and N$_2$
containing CO ices, and, less likely, processed O$_2$--rich CO ices
could be the constituents of apolar ices.

\subsection{Tracing Chemical Pathways through Isotope Ratios}

In this work, we found that the solid \twelveco/\thirteenco\ abundance
ratio is 71$\pm$15 (3$\sigma$) in the \irsnine\ line of sight (\S
3.2).  This value is in good agreement with the gas phase CO isotope
ratio of 78 that is expected at the galactocentric radius of \irsnine\
(9.4 kpc; \citealt{wils94}).  Furthermore, the isotope ratios for gas
and solid CO are similar to that found for solid state
\twelveco$_2$/\thirteenco$_2$ along the same line of sight
(80$\pm$11\footnote{The error quoted for the
\twelveco$_2$/\thirteenco$_2$ ratio represents a systematic error, the
3$\sigma$ statistical error is less than this.}; \citealt{boog00}).
These values are expected to trace the `true' carbon isotope ratio at
this galactocentric radius (see below), which provides constraints on
models for the chemical evolution of the Galaxy \citep{tosi82}.  Also,
the isotope ratios provide an independent test for the chemical origin
of interstellar CO$_2$.

Chemical fractionation, or isotope selective destruction, expressed in
the isotopic exchange reaction
\begin{equation}
~^{13}{\rm C}^+ + ~^{12}{\rm CO} \leftrightarrows ~^{13}{\rm CO} +
~^{12}{\rm C}^+ +35\ {\rm K,}~\label{eq1}
\end{equation}
can greatly change the atomic $^{12}{\rm C}^{(+)}/^{13}{\rm C}^{(+)}$
ratio even deep within dense clouds.  Species derived from atomic
C$^{\rm (+)}$ will reflect the enhanced isotope ratio.  On the other
hand, the isotope ratio of CO, and its chemical daughter products, is
not significantly affected by these processes in dense clouds, where
CO is the main carbon reservoir.  Within this framework, the carbon
isotope ratio of a variety of molecules in dense clouds has been
investigated in chemical models \citep{lang89}.  These models show
isotope ratios of 120--220 for H$_2$CO, CS, and HCN compared to 73 for
CO. The wide range of values reflects the range of temperatures and
densities used in the models. The ratios increase with temperature and
density. In the case of \irsnine, the large abundance of apolar CO
ices indicates dust temperatures less than the sublimation temperature
($< 18$ K; \citealt{naka80}). This would give carbon isotope ratios of
trace species of 120-150 at a density of $5\times 10^3$ \cubcm.  Such
densities are realistic in the cold outer parts of the envelope around
\irsnine\ \citep{tak00}. We conclude that while the observed solid
\twelveco/\thirteenco\ ratio is comparable to that of the models, gas
phase models in which CO$_2$ is formed from atomic $\rm C^{(+)}$ would
predict a significantly higher ratio for CO$_2$. The similarity of the
isotope ratios for solid CO and CO$_2$ thus indicates, not
unexpectedly, that interstellar CO$_2$ has been formed from CO.  This
is consistent with proposed grain surface chemistry schemes based upon
oxidation of accreted CO \citep{tiel82}. The absence of CO$_2$ in the
main, 2140 \waven, apolar CO component of \irsnine\ (\S 3.1.1)
indicates that CO$_2$ was formed at a different time or location than
the apolar CO ice mantles.  Indeed, CO$_2$ was found to be intimately
mixed with H$_2$O \citep{gera99}, itself a grain surface product.  The
similarity of the solid CO and CO$_2$ isotope ratios is also
consistent with energetic processing (UV irradiation, cosmic ray hits)
of existing CO ice mantles \citep{dhen86}, but this is {\it not}
favored by the narrow width of the 2140 \waven\ apolar CO component.

While there is little doubt that CO$_2$ originates from CO, the above
discussion illustrates that isotope ratios may provide a powerful way
to discriminate between gas phase and grain surface routes for
molecules of which the origin is more disputed.  In particular,
CH$_3$OH and H$_2$CO are often assumed to reflect evaporation of ices,
even in dark clouds (e.g. \citealt{cecc01}).  Their formation on the
grain surface probably involved hydrogenation of CO
(e.g. \citealt{char92}).  Their isotope ratio should thus reflect that
of CO.  In contrast, gas phase models for the formation of these
species rely on C$^+$ or C broken out of CO through, for example,
cosmic ray produced photons.  Their isotope ratio is then expected to
be quite different \citep{lang89}. Future accurate determinations of
carbon ratios of gas phase H$_2$CO and CH$_3$OH and more complex
species will therefore be useful in unraveling the interstellar
chemical network.

\acknowledgments

The research of A.C.A.B. and G.A.B. is supported by the SIRTF Legacy
Science program and by the Owens Valley Radio Observatory through NSF
grant AST-9981546. We thank the anonymous referee for several useful
comments. The authors wish to extend special thanks to those of
Hawaiian ancestry on whose sacred mountain we are privileged to be
guests.  Without their generous hospitality, none of the observations
presented herein would have been possible.

\end{document}